\documentclass[aps,prd,twocolumn,superscriptaddress,showpacs]{revtex4-2}
\usepackage{graphicx}
\usepackage{amsmath}
\usepackage{amsfonts}
\usepackage{amssymb}
\usepackage{amssymb}
\usepackage{slashed}
\usepackage{pstricks,pst-coil}
\usepackage{dsfont}
\usepackage{bm}

\begin{document}
\title{Classical features, Anderson-Higgs mechanism, and unitarity in Lee-Wick pseudo-electrodynamics}
%
%


\author{M. J. Neves}\email{mariojr@ufrrj.br}

\affiliation{Departamento de F\'isica, Universidade Federal Rural do Rio de Janeiro, BR 465-07, 23890-971, Serop\'edica, Rio de Janeiro, Brazil}

\begin{abstract}

In this paper, the dimensional reduction is applied to the Lee-Wick electrodynamics in which the classical sources are confined on a spatial plane.
As result, the Lee-Wick pseudo-electrodynamics is achieved as a non-local electromagnetism defined in $1+2$ dimensions. The abelian Anderson-Higgs 
mechanism is so introduced in the Lee-Wick pseudo-electrodynamics through a complex scalar sector in $1+2$ dimensions, breaking spontaneously the $U(1)$-gauge 
symmetry of the non-local theory. As consequence, the pseudo-Lee-Wick field acquires a light mass, beyond the usual heavy Lee-Wick mass, that is a natural mass 
parameter of the theory. After the spontaneous symmetry breaking takes place, classical features of the theory are discussed, as the Proca-Lee-Wick pseudo-electrodynamics, 
with the field equations and conservation laws. The introduction of Lee-Wick fermions also is proposed, in which it opens the discussion of a viable 
Proca-Lee-Wick pseudo-quantum electrodynamics in $1+2$ dimensions. The unitarity at the tree level of the Lee-Wick pseudo-electrodynamics is discussed 
through the Optical theorem.  
%

%
%

\end{abstract}
\maketitle

%
%



\section{Introduction}

%
%
Quantum field theories (QFTs) in two spatial dimensions have gained relevance with several applications in
condensed matter physics in the last years. As examples, the quantum Hall effect \cite{Ando,Tsui,Laughlin,Chamon}, 
topological planar materials \cite{Qi,Hasan,Chiu,Zhao,Qi2008}, the study of transport in graphene \cite{Gorbar1,Herbut,Gusynin,Herbut2}, 
superconductivity in layered materials \cite{Tesanovic,Zhang,Franz,Kivelson,Marino2018}, and among others show that field theories in low 
dimensions are excellent theoretical approach to explain experimental results in material physics. In special, Dirac materials are systems 
that exhibit a semi-relativistic dynamics for massless or massive particles with Fermi velocity and that are good tests for QFTs \cite{Castro}. 
The interaction electron-electron in planar materials must so be mediated by a gauge field defined in $1+2$ dimensions, following the basic 
principle from quantum electrodynamics (QED). The gauge theory that describe the interactions of fermions in a planar QED is known 
as pseudo-electrodynamics (PED) \cite{Marino93}. Basically, the PED is obtained from usual Maxwell ED when the classical sources of 
charges and currents are confined on a spatial plane, reducing the space-time from $1+3$ to $1+2$ dimensions 
in the generating functional of the Abelian gauge theory. As consequence of the reduced dimension, the PED is a non-local ED with derivatives 
of infinity order in the D'Alembertian operator. The PED preserves the gauge symmetry and the continuity equation for the current. As fundamental ingredients 
of a QFT, the PED also preserves the properties of causality \cite{Amaral}, and unitarity \cite{Marino2014}. Extensions of the PED has been investigated 
in the literature, as the addition of topological Chern-Simons term \cite{Alves,Alves2}, the dimensional reduction of the Proca ED
that leads to Pseudo-Proca ED \cite{Ozela,Ozela2,VanSergio}, and the dimensional reduced QED applied to planes and fermion gap generation \cite{Gorbar}. 
In brane physics, the dynamical chiral symmetry breaking was studied through the dimensional reduction of the QED \cite{GorbarPRD}. The dimensional reduction 
in connection with boundary conditions were studied for scalar and abelian gauge theories \cite{Edery}.
However, the Proca theory breaks the gauge invariance, and others extensions of the Maxwell, that preserve the gauge invariance can be the good source of 
investigation to include a massive gauge field in planar materials. One of these possibilities is in the Lee-Wick ED \cite{lw69,lw70}, that also is known as 
Podolsky ED \cite{podolsky42,podolsky44}. The Lee-Wick theory is a ED with higher derivative in the kinetic term, that introduces naturally a massive 
degree of freedom called Lee-Wick mass, preserves the gauge invariance. The Lee-Wick static potential is the subtraction of Coulombian by the Yukawa potential, 
and as consequence, it is finite at the origin. The Maxwell ED is recovered when the Lee-Wick mass is infinity, and induces the 
idea that the particle associated with the Lee-Wick field is heavy. From point of view of QFT, the motivation to study the Lee-Wick ED is in the fact 
of the propagator to have a better behaviour in the ultraviolet regime, and that helps in the renormalizability of the theory in 4D. Therefore, 
many investigations were studied in literature \cite{accioly03,accioly04,accioly05,accioly11,turcati14,turcati15}, 
including the construction of a Standard Model for elementary particles based on Lee-Wick approach \cite{Grinstein}. 
Thus, in low dimensions, the Lee-Wick ED can be a super-renormalizable or finite in the first orders of the perturbation theory. 
This is the main motivation to investigate the aspects of the Lee-Wick ED when applied to the planar materials.          
In this paper, we show the dimensional reduction to $1+2$ dimensions 
in the Lee-Wick ED, where the sources are constraint on a spatial plane. 
Thereby, the non-local theory called Lee-Wick pseudo-electrodynamics is obtained 
and defined in two spatial dimensions and one time-coordinate \cite{TesePedro}. The known pseudo-ED is 
so recovered when the Lee-Wick mass (that is a natural mass parameter of the theory) 
goes to infinity. Motivated by the Abelian Chern-Simons model applied to planar superconductors \cite{Xing}, 
we propose a Higgs-Anderson mechanism to break spontaneously the $U(1)$-gauge symmetry via a complex scalar 
field in $1+2$ dimensions, whose vacuum expected value yields a mass to the Lee-Wick pseudo-electrodynamics, 
beyond the Lee-Wick mass parameter. For simplification, it is considered that the mass acquired via 
spontaneous symmetry breaking (SSB) is lighter in relation to Lee-Wick mass. The scalar sector after the SSB 
also is showed as a toy model that is super-renormalizable in $1+2$-dimensions. It is discussed some 
classical features of the Proca-Lee-Wick pseudo-ED, as the field equations and the correspondent conservations 
laws. Posteriorly, the fermion sector is introduced as a starting point for discussion of the viable 
pseudo-quantum electrodynamics in the presence of a Proca mass for the Lee-Wick field in $1+2$ dimensions.
As important consistency of a quantum field theory, the conditions for the unitarity of the Lee-Wick 
pseudo-electrodynamics are showed through the Optical theorem \cite{Peskin}.          
The paper is organized as follows : In the section \ref{sec2}, the dimensional reduction of the Lee-Wick 
is showed to yield the Lee-Wick pseudo-ED. The section \ref{sec3} is dedicated to the Anderson-Higgs mechanism
applied to Lee-Wick pseudo-ED. Some classical properties of the Proca-Lee-Wick pseudo-ED are showed in the section \ref{sec4}.
In the section \ref{sec5} is proposed the fermion sector coupled to the Proca-Lee-Wick field in $1+2$ dimensions.
The section \ref{sec6} is dedicated to discussion of the unitarity in Lee-Wick pseudo-ED at three level. For end, 
the conclusions are highlighted in the section \ref{sec7}.
The natural units system $\hbar=c=1$ are used throughout the paper. The signature of the metric is $\eta_{\mu\nu}=\mbox{diag}(+1,-1,-1,-1)$ 
for the theory in $1+3$ dimensions. In the theory dimensionally reduced to $1+2$ dimensions, we use the bar index $\bar{\mu}=\{0,1,2\}$ 
for vectors and tensors, with the metric $\eta_{\bar{\mu}\bar{\nu}}=\mbox{diag}(+1,-1,-1)$.


\section{The dimensional reduction in Lee-Wick electrodynamics}
\label{sec2}

The well-known Lee-Wick ED is set by the lagrangian density in the presence of a gauge fixing term
\begin{eqnarray} \label{LLW}
\mathcal{L}_{LW} &=& -\frac{1}{4} \, F_{\mu\nu}\left(1+\frac{\Box}{M^2}\right)F^{\mu\nu} 
\nonumber \\
&&
-\frac{1}{2\xi} \, \left[  \left(1+\frac{\Box}{M^2} \right) \partial_{\mu}A^{\mu} \right]^2 
- J_{\mu} \, A^{\mu}  \; ,
\end{eqnarray}
where $F_{\mu\nu}=\partial_{\mu}A_{\nu}-\partial_{\nu}A_{\mu}$ is the strength field tensor of the $A^{\mu}$-potential, $M$ is the Lee-Wick mass for $A^{\mu}$, $\xi$ is a real gauge fixing parameter, and $J^{\mu}$ is a external classical source. The limit $M \rightarrow \infty$ recovers the usual Maxwell ED with a covariant gauge fixing term. Thereby, the Lee-Wick mass represents degree freedom of a heavy gauge boson. The action principle applied to the lagrangian (\ref{LLW}) yields the field equation for the $A^{\mu}$-potential
\begin{eqnarray}\label{EqOpA}
\mathcal{O}_{\mu\nu} \, A^{\mu} = J_{\nu} \; ,
\end{eqnarray}
in which $\mathcal{O}_{\mu\nu}$ is the operator
\begin{eqnarray}\label{Op}
\mathcal{O}_{\mu\nu}=\left( 1+\frac{\Box}{M^2} \right) \left(\, \eta_{\mu\nu} \, \Box-\partial_{\mu}\,\partial_{\nu} \, \right) 
\nonumber \\
+ \frac{1}{\xi} \left( 1+\frac{\Box}{M^2} \right)^2 \! \partial_{\mu} \, \partial_{\nu} \; .
\end{eqnarray}
The solution for the equation (\ref{EqOpA}) is expressed by 
\begin{eqnarray}\label{SolA}
A_{\mu}(x)=A_{\mu}^{(0)}(x)+\int d^4 x^{\prime} \, \Delta_{\mu\nu}(x-x^{\prime}) \, J^{\nu}(x^{\prime}) \; , 
\end{eqnarray}
where $A_{\mu}^{(0)}$ is the solution of the homogeneous equation $\mathcal{O}_{\mu\nu}A^{(0)\mu}=0$, and $\Delta_{\mu\nu}(x-x^{\prime})$ is the 
Green function associated with the operator (\ref{Op}) that satisfies the equation 
\begin{eqnarray}\label{EqGreen}
\mathcal{O}_{\mu\nu} \, \Delta^{\nu}_{\;\, \alpha}(x-x^{\prime})= \delta^{\mu}_{\;\,\, \alpha} \, \delta^{4}(x-x^{\prime}) \; .
\end{eqnarray}
After integrations by parts, the functional action can be written as  
\begin{eqnarray}
S[A_{\mu}]=\int d^4x \, \left[ \, \frac{1}{2} \, A_{\mu} \, \mathcal{O}^{\mu\nu} \, A_{\nu} - J_{\mu}\,A^{\mu} \, \right] \; ,
\end{eqnarray}
that substituting the solution (\ref{SolA}), it is reduced to the functional of the external source    
 \begin{equation}
S[J_{\mu}]=-\frac{1}{2} \int d^4x \, d^4x^{\prime} \, J_{\mu}(x) \, \Delta^{\mu\nu}(x-x^{\prime}) \, J_{\nu}(x^{\prime}) \; . 
\end{equation}
For convenience, the dimensional reduction is realized with the action written in the euclidian $4D$ space, 
with $x^0=-i\,x_4$ and $\Box \rightarrow -\Box_{E}$,  in which
%
%
the operator (\ref{Op}) is given by
\begin{equation}\label{OpE}
\mathcal{O}_{E}^{\mu\nu}=\left[ 1+\frac{(-\Box_{E})}{M^2} \right](-\Box_{E}) \left\{ \theta_{E}^{\mu\nu}
+\frac{1}{\xi}\left[ 1+\frac{(-\Box_{E})}{M^2} \right] \omega_{E}^{\mu\nu} \right\} \; .
\end{equation}
The projectors in the euclidian space are defined by
\begin{equation}
\theta_{E}^{\mu\nu}= \delta_{\mu\nu}-\frac{\partial_{E}^{\mu} \, \partial_{E}^{\nu} }{(-\Box_{E})}
\quad , \quad
\omega_{E}^{\mu\nu}=\frac{\partial_{E}^{\mu} \, \partial_{E}^{\nu} }{(-\Box_{E})} \; ,
\end{equation}
that satisfy the relations
\begin{equation}
\theta_{E}^{ \; \; \, \mu\alpha} \, \theta_{E \, \alpha\nu} = \theta_{E \; \; \nu}^{ \;\;\; \mu} 
\; , \; 
\omega_{E}^{ \; \; \, \mu\alpha} \, \omega_{E \, \alpha\nu}=\omega_{E \; \; \nu}^{ \;\;\; \mu}
\; , \;
\theta_{E}^{ \; \; \, \mu\alpha} \, \omega_{E \, \alpha\nu} = 0 \; .
\end{equation}
Using the Fourier transform in the momentum space (euclidian), the Green function that satisfies (\ref{EqGreen}) is read below 
\begin{eqnarray}
\Delta_{\mu\nu}(x_E-x_E^{\prime})=\int \frac{d^4k_{E}}{(2\pi)^4} \frac{M^2}{k_{E}^2(k_{E}^2+M^2)} \, \times
\nonumber \\
\times \, \left[ \, \delta_{\mu\nu}-\frac{ k^{E}_{\mu} \, k^{E}_{\nu} }{k_{E}^2} \frac{ k_{E}^2+M^2(1-\xi)}{ k_{E}^2+M^2} \, \right] 
e^{ik_{E}\cdot(x_E-x_E^{\prime})}
\; , \;\;\;\;
\end{eqnarray}
where $k_{E}^{\mu}=(k_{4}=i\,k_{0},k_{x},k_{y},k_{z})$ is the $4$-momentum in the euclidian space, and $k_{E}^2=k_{4}^2+k_x^2+k_{y}^2+k_{z}^{2}$ is positive.
Consequently, the effective action in the euclidian space is 
\begin{equation}\label{actionEJ}
S_{E}[J_{\mu}]=-\frac{1}{2} \int d^4x_E\,d^4x_E^{\prime} \, J_{\mu}(x_E)\,\Delta^{\mu\nu}(x_E-x_E^{\prime}) \, J_{\nu}(x_E^{\prime}) \; . 
\end{equation}
The dimensional reduction is so introduced constraint the external sources on the 2D spatial plane    
\begin{subequations}
\begin{eqnarray}
J^{\bar{\mu}}(x_{E}^{\mu}) &=& j^{\bar{\mu}}(x_E^{\bar{\mu}})\,\delta(z) \; ,
\label{Jmu}
\\
J^{3}(x^{\mu}) &=& 0 \; ,
\label{J3}
\end{eqnarray}
\end{subequations}
where $x_{E}^{\bar{\mu}}=(x_4,x,y)$ sets the coordinates on the 3D euclidian space-time.
Substituting these conditions in (\ref{actionEJ}), the euclidian action is reduced to a 3D space-time as 
%
%
\begin{equation}\label{actionEJ3D}
S_{E}[j_{\bar{\mu}}]=-\frac{1}{2} \int d^3\bar{x}_E\,d^3\bar{x}_E^{\prime} \, j_{\bar{\mu}}(\bar{x}_E)\,\Delta^{\bar{\mu}\bar{\nu}}(\bar{x}_E-\bar{x}_E^{\prime}) \, j_{\bar{\nu}}(\bar{x}_E^{\prime}) \; , 
\end{equation}
where the new Green function in 3D is 
\begin{eqnarray}\label{Deltamunu}
&&
\left.
\Delta_{\bar{\mu}\bar{\nu}}(\bar{x}_{E}-\bar{x}_{E}^{\prime})=\Delta_{\bar{\mu}\bar{\nu}}(x_{E}-x_{E}^{\prime}) \right|_{z=z^{\prime}=0}
\nonumber \\
&&
\left.
=\int \frac{d^4k_{E}}{(2\pi)^4} 
\frac{M^2 \, \delta_{\bar{\mu}\bar{\nu}} }{k_{E}^2(k_{E}^2+M^2)}
\,  e^{ik_{E}\cdot(x_E-x_E^{\prime})} \right|_{z=z^{\prime}=0} \, ,
\hspace{0.5cm}
\end{eqnarray}
in which we have used the conserved current $k_{\bar{\mu}} \, j^{\bar{\mu}}=0$ in the momentum space. 
The $k_z$-integration yields the result
\begin{eqnarray}\label{Deltamunuresult}
\Delta_{\bar{\mu}\bar{\nu}}(\bar{x}_{E}-\bar{x}_{E}^{\prime})
=\frac{\delta_{\bar{\mu}\bar{\nu}}}{2} \int \frac{d^3\bar{k}_{E}}{(2\pi)^3} \, e^{i \bar{k}_{E} \cdot ( \bar{x}_{E}-\bar{x}_{E}^{\prime}  ) } 
\nonumber \\
\times \left[ \frac{1}{\sqrt{\bar{k}_{E}^2}}-\frac{1}{\sqrt{\bar{k}_{E}^2+M^2}} \right] \; , \;\;\;\;
\end{eqnarray}
where $\bar{k}_{E}^2=k_{E\bar{\mu}}k_{E}^{\bar{\mu}}=k_{4}^2+k_{x}^2+k_{y}^2>0$. Calculating the integrals in (\ref{Deltamunuresult}), 
the Green function in the euclidian space is 
\begin{eqnarray}
\Delta_{\bar{\mu}\bar{\nu}}(\bar{x}_{E}-\bar{x}_{E}^{\prime})
=\frac{\delta_{\bar{\mu}\bar{\nu}}}{4\pi^2|\bar{x}_{E}-\bar{x}_{E}^{\prime}|^2} 
\nonumber \\
\times
\left[ \, 1- M|\bar{x}_{E}-\bar{x}_{E}^{\prime}|\,K_{1}(M|\bar{x}_{E}-\bar{x}_{E}^{\prime}|) \, \right] \; ,
\end{eqnarray}
where $K_{1}$ is a Bessel function of second kind, and the effective action (\ref{actionEJ3D}) is
\begin{eqnarray}
S_{E}[j_{\bar{\mu}}]=-\frac{1}{8\pi^2} \int d^3\bar{x}_E\,d^3\bar{x}_E^{\prime} \,
\frac{j_{\bar{\mu}}(\bar{x}_E) \, j^{\bar{\mu}}(\bar{x}_E^{\prime})}{|\bar{x}_E-\bar{x}_E^{\prime}|^2}
\nonumber \\
\times \left[ \, 1- M|\bar{x}_{E}-\bar{x}_{E}^{\prime}|\,K_{1}(M|\bar{x}_{E}-\bar{x}_{E}^{\prime}|) \, \right]
 \; .
\end{eqnarray}
Thereby, we seek now the Lagrangian density defined in the $1+2$ space-time that leads to the Green function (\ref{Deltamunuresult}).   
The proposed Lagrangian must be in the form
\begin{eqnarray} \label{PLLW}
\mathcal{L}_{PLW} &=& -\frac{1}{4} \, F_{\bar{\mu}\bar{\nu}} \left(1+\frac{\bar{\Box}}{M^2}\right)W(\bar{\Box})F^{\bar{\mu}\bar{\nu}} 
\nonumber \\
&&
+\frac{1}{2\xi} \, A_{\bar{\mu}}  \left(1+\frac{\bar{\Box}}{M^2} \right)^2 W(\bar{\Box}) \, \partial^{\bar{\mu}}\,\partial^{\bar{\nu}} A_{\bar{\nu}} 
\nonumber \\
&&
- j_{\bar{\mu}} \, A^{\bar{\mu}}  \; ,
\end{eqnarray}
where $\bar{\Box}=\partial_{\bar{\mu}}\partial^{\bar{\mu}}=\partial_{t}^2-\partial_{x}^2-\partial_{y}^2$ is the D'Alembertian operator in $1+2$ dimensions,
$W(\bar{\Box})$ is a operator that depends on $\bar{\Box}$,  $F_{\bar{\mu}\bar{\nu}}=\partial_{\bar{\mu}}A_{\bar{\nu}}-\partial_{\bar{\nu}}A_{\bar{\mu}}$
is the EM tensor defined on the $1+2$ space-time, and $A^{\bar{\mu}}=(A^{0},A_{x},A_{y})$ is the correspondent potential. The space-time derivatives
$\partial_{\bar{\mu}}=(\partial_{t},\partial_{x},\partial_{y})$ acts on the fields of the theory. If we repeat the same steps for the Lagrangian (\ref{PLLW}), 
the correspondent Green function in $1+2$ is given by  
\begin{eqnarray}\label{DeltaWk}
\Delta_{\bar{\mu}\bar{\nu}}(\bar{x}-\bar{x}^{\prime})
=\int \frac{d^3\bar{k}}{(2\pi)^3}\frac{M^2}{W(-\bar{k}^2)\bar{k}^2(\bar{k}^2-M^2)} 
\nonumber \\
\times \left[ \, \theta_{\bar{\mu}\bar{\nu}}-\frac{M^2\,\xi}{\bar{k}^2-M^2} \, \omega_{\bar{\mu}\bar{\nu}} \, \right] e^{i\bar{k}\cdot(\bar{x}-\bar{x}^{\prime})} \; ,
\end{eqnarray}
and comparing it with (\ref{Deltamunuresult}), the $W(-\bar{k}^2)$-function is 
\begin{equation}\label{Wk}
W(-\bar{k}^2)=\frac{2\,M^2}{(-\bar{k}^2+M^2)\sqrt{-\bar{k}^2}-(-\bar{k}^2)\sqrt{-\bar{k}^2+M^2} } \; .
\end{equation}
Using the representation of $\bar{k}^2$ as $(-\bar{\Box})$, the $W(\bar{\Box})$-operator is
\begin{eqnarray}\label{Wop}
W(\bar{\Box})=\frac{2M^2}{(\bar{\Box}+M^2)\sqrt{\bar{\Box}}-\bar{\Box}\,\sqrt{\bar{\Box}+M^2}} \; , 
\end{eqnarray}
and the lagrangian (\ref{PLLW}) can be written as
\begin{eqnarray} \label{PLLWresultOpN}
\mathcal{L}_{PLW} &=& -\frac{1}{4} \, F_{\bar{\mu}\bar{\nu}}N(\bar{\Box})F^{\bar{\mu}\bar{\nu}} 
\nonumber \\
&&
+\frac{1}{2\xi} \, A_{\bar{\mu}}  \left(1+\frac{\bar{\Box}}{M^2} \right) N(\bar{\Box}) \, \partial^{\bar{\mu}}\,\partial^{\bar{\nu}} A_{\bar{\nu}} 
\nonumber \\
&&
- j_{\bar{\mu}} \, A^{\bar{\mu}}  \; ,
\end{eqnarray}
where we have defined the $N(\bar{\Box})$-operador as
\begin{eqnarray}\label{Nop}
N(\bar{\Box})=\frac{2(\bar{\Box}+M^2)}{(\bar{\Box}+M^2)\sqrt{\bar{\Box}}-\bar{\Box}\,\sqrt{\bar{\Box}+M^2}} \; .
\end{eqnarray}
The kinetic term in (\ref{PLLWresultOpN}) can be written in form of field-operator-field as
\begin{equation}
{\cal L}_{PLW}^{K}=A_{\bar{\mu}}\left[\left(1+\frac{\bar{\Box}}{M^2} \right)\sqrt{\bar{\Box}}
+\frac{\bar{\Box}}{M^2}\sqrt{1+\frac{\bar{\Box}}{M^2}} \right] \theta^{\bar{\mu}\bar{\nu}}A_{\bar{\nu}} \; ,
\end{equation}
%
%
and show that the dimensional reduction leads to a non-local electrodynamics of infinity order in the D'Alembertian operator $\bar{\Box}$. 
The theory (\ref{PLLWresultOpN}) can be called pseudo-Lee-Wick electrodynamics, in which the limit $M \rightarrow \infty$ 
recovers the pseudo-electrodynamics discussed in the refs. \cite{Marino93,Marino2014}. The lagrangian (\ref{PLLWresultOpN}) (up to gauge fixing term)
is $U(1)$ gauge invariant under the transformation $A_{\bar{\mu}} \mapsto A^{\prime}_{\bar{\mu}}=A_{\bar{\mu}}-\partial_{\bar{\mu}}\Lambda$, if the current 
density is conserved, {\it i.e.}, it obeys the continuity equation $\partial_{\bar{\mu}}j^{\bar{\mu}}=0$ on the plane. Substituting the result of the 
$W(-\bar{k}^2)$-function (\ref{Wk}) in (\ref{DeltaWk}), the Green function in the $\xi$-gauge fixing is 
\begin{eqnarray}\label{DeltamunuPLWresult}
\Delta_{\bar{\mu}\bar{\nu}}(\bar{x}-\bar{x}^{\prime})
=\int \frac{d^3\bar{k}}{(2\pi)^3} \left[ \frac{1}{2\sqrt{-\bar{k}^2}}-\frac{1}{2\sqrt{-\bar{k}^2+M^2}} \right]
\nonumber \\
\times \left[ \, \eta_{\bar{\mu}\bar{\nu}}-\frac{\bar{k}^2+M^2(\xi-1)}{\bar{k}^2-M^2} \, \frac{k_{\bar{\mu}} \, k_{\bar{\nu}} }{\bar{k}^2} \, \right] e^{i\bar{k}\cdot(\bar{x}-\bar{x}^{\prime})} \; . \;\;\;\;
\end{eqnarray}
Notice that the expression Green function in the momentum space has two poles : $\bar{k}^2=0$ and $\bar{k}^2=M^2$. Thus, the theory shows one massless 
degree freedom, and another massive degree freedom described by the Lee-Wick mass. The case of two like-point charged particles $(-e)$ at rest, the current density 
is $j^{\bar{\mu}}({\bf r})= [ -e\,\delta^{3}({\bf r}-{\bf r}_1)-e\,\delta^{3}({\bf r}-{\bf r}_2) \, , \, {\bf 0} ] $, and the effective action reduces to energy of interaction between these 
two particles :  
\begin{equation}
U(r)=-e^2\int \frac{d^2{\bf k}}{(2\pi)^2} \left[ \frac{e^{i {\bf k}\cdot{\bf r}}}{|{\bf k}|} - \frac{e^{i {\bf k}\cdot{\bf r}}}{\sqrt{{\bf k}^2+M^2}} \right] \; ,
\end{equation}
whose result is well known in the literature
\begin{eqnarray}\label{UrLW}
U(r)=-\frac{e^2}{4\pi r} \, \left( \, 1-e^{-Mr} \, \right) \; ,
\end{eqnarray}
where $r=|{\bf r}_{1}-{\bf r}_{2}|$ is the distance that separate the charges. The energy (\ref{UrLW}) is the same of the usual 
Lee-Wick ED that is the difference of the Coulomb by the Yukawa potential with the Lee-Wick mass.
It is worth to clarify that the Pseudo-Lee-Wick ED is a causal theory. As obtained in the ref. \cite{TesePedro}, 
the retarded and advanced Green functions associated with the equation (\ref{EqOpA}) when $\xi \rightarrow \infty$ 
in the prescription of $k_{4}\rightarrow k_{0} \pm i \epsilon$ are given by
\begin{eqnarray}
\Delta^{\pm}(\bar{x}-\bar{x}^{\prime})=-\frac{i}{\pi} \, \Theta(\pm\tau) \left\{ \, \delta[(\bar{x}-\bar{x}^{\prime})^2]
\right. 
\nonumber \\
\left.
-\frac{1}{2}\,\Theta[(\bar{x}-\bar{x}^{\prime})^2]\, \frac{M\, J_{1}(M\,\sqrt{(\bar{x}-\bar{x}^{\prime})^2})}{\sqrt{(\bar{x}-\bar{x}^{\prime})^2}} \right\} \; ,
\end{eqnarray}
where the signs $(\pm)$ indicate the retarded $(+)$ and advanced $(-)$, $\tau=t-t^{\prime}$, $\Theta$ is the Heaviside function, and $J_{1}$ is Bessel function 
of first kind. The result shows that the retarded Green function is null outside the light-cone, for $(\bar{x}-\bar{x}^{\prime})^2<0$. 
The Feynman Green function is obtained with the prescription of $k_{4}^2 \rightarrow k_{0}^2+i\epsilon$, such that the result is \cite{TesePedro}
\begin{eqnarray}
\Delta_{F}(\bar{x}-\bar{x}^{\prime})=-\frac{1}{2\pi^2}\left[ \, \frac{1}{(\bar{x}-\bar{x}^{\prime})^2+i\epsilon}
\right. 
\nonumber \\
\left.
-\frac{i\pi}{2} \, \frac{M\,H_{1}(-M\sqrt{(\bar{x}-\bar{x}^{\prime})^2+i\epsilon})}{\sqrt{ (\bar{x}-\bar{x}^{\prime})^2+i\epsilon } } \, \right] \; ,
\end{eqnarray}
in which $H_{1}$ is a Hankel function of first kind. 
\section{The Anderson–Higgs mechanism in the pseudo-Lee Wick ED}
\label{sec3}
The sector of a complex scalar field $\phi(x^{\bar{\mu}})$ $U(1)$-gauge invariant is governed by the Lagrangian  
\begin{eqnarray} \label{lwl}
\mathcal{L}_{sc} &=& 
|D_{\bar{\mu}}\phi|^{2}-V(\phi^{\ast},\phi)
+\frac{1}{M_{\phi}^2}|D_{\bar{\mu}}D^{\bar{\mu}}\phi|^2 \; ,
\end{eqnarray}
where 
$V(\phi^{\ast},\phi)=\mu^2|\phi|^2+\lambda|\phi|^4$ is the scalar potential, $\mu^2$ and $\lambda$ are two real parameters, and $M_{\phi}$ is the Lee-Wick mass for the scalar field.  The covariant derivative operator is $D_{\bar{\mu}}\phi=\partial_{\bar{\mu}}\phi+igA_{\bar{\mu}}\phi$, in which $g$ is a real coupling constant. The complex scalar field $\phi$ under a local transformation keeps the lagrangian (\ref{lwl}) $U(1)$-gauge invariant in $1+2$ dimensions. As in the usual mechanism, the scalar potential acquires the vacuum expected value (VEV) $v\equiv|\phi_{0}|=\sqrt{-\mu^2/(2\lambda)}$, with $\mu^2<0$, and therefore, the scalar field can be written as the perturbation in the unitary gauge
\begin{eqnarray}\label{phih}
\phi(\bar{x})=\frac{\sqrt{v}+\hat{h}(\bar{x})}{\sqrt{2}} \; .
\end{eqnarray}
Notice that $\phi$ has dimension of mass elevated to $1/2$ in three dimensions, such that the combination (\ref{phih}) is dimensionally satisfied in 3D.  
After the SSB mechanism, the free sector of (\ref{lwl}) is
\begin{equation}\label{Leff0}
{\cal L}_{sc}^{(0)}=
\frac{1}{2}\, (g^2v) \,A_{\bar{\mu}}^2
+\frac{1}{2}\,(\partial_{\bar{\mu}}\hat{h})^2-\frac{1}{2}\,m_{h}^2\,\hat{h}^2+\frac{1}{2M_{\phi}^2}\,(\bar{\Box}\hat{h})^2 \; ,
\end{equation}
where the term $g^2\,v$ is a mass acquired by the gauge field $A^{\mu}$ due to VEV scale, and $m_{h}=\sqrt{2\lambda\, v^2}$ is a mass acquired by the scalar field 
$\hat{h}$. We consider in this paper the condition in that $M,M_{\phi} \gg v$, and consequently, the Lee-Wick masses $M$ and $M_{\phi}$ are heavier in relation to masses 
$m_{a}=g^2\,v$ and $m_{h}$. 
The pure scalar sector of $\hat{h}$-field is read below
\begin{equation}\label{Leffh}
{\cal L}_{sc}^{(\hat{h})}=\frac{1}{2}\,\partial_{\bar{\mu}}\hat{h}\left(1+\frac{\bar{\Box}}{M_{\phi}^2} \right)\partial^{\bar{\mu}}\hat{h}
-\frac{1}{2} \, m_{h}^2 \, \hat{h}^2
-\lambda\,\sqrt{v}\,\hat{h}^3-\frac{\lambda}{4}\,\hat{h}^4 \; ,
\end{equation}
that sets a toy model in 3D with self-interactions of $\hat{h}^3$ and $\hat{h}^4$ of coupling constants 
$\lambda\,\sqrt{v}$ and $\lambda$, respectively. The free sector in 
(\ref{Leffh}) yields the scalar propagator in the momentum space $1+2$
\begin{eqnarray}\label{propPhi}
\Delta(\bar{p}) &=& \frac{i}{\bar{p}^2\left(1-\frac{\bar{p}^2}{M_{\phi}^2}\right)-m_{h}^2}
\nonumber \\
&&
\hspace{-0.5cm}
=\frac{M_{\phi}}{\sqrt{M_{\phi}^2-4m_h^2}}\left[ \, \frac{i}{\bar{p}^2-\mu_1^2}-\frac{i}{\bar{p}^2-\mu_2^2} \, \right] \; ,                                                  
\end{eqnarray}
that constraints the condition $M_{\phi} > 2\, m_{h}$, and $\mu_{1}$, $\mu_{2}$ are the mass eigenstates :
\begin{subequations}
\begin{eqnarray}
\mu_{1}^2 &=& \frac{M_{\phi}^2}{2}-\frac{M_{\phi}^2}{2}\sqrt{1-\frac{4\,m_{h}^2}{M_{\phi}^2}} \; ,
\\
\mu_{2}^2 &=& \frac{M_{\phi}^2}{2}+\frac{M_{\phi}^2}{2}\sqrt{1-\frac{4\,m_{h}^2}{M_{\phi}^2}} \; .
\end{eqnarray}
\end{subequations}
In the case of $M_{\phi} \gg m_h$, these eigenstates are reduced to $\mu_{1} \simeq m_{h}$ and $\mu_{2}\simeq M_{\phi}$, 
respectively, and simplifies (\ref{propPhi}) as
\begin{equation}\label{propPhisimpl}
\Delta(\bar{p})\simeq \frac{i}{\bar{p}^2-m_h^2}-\frac{i}{\bar{p}^2-M_\phi^2} \; ,                                                  
\end{equation} 
which it is clearly the composition of a scalar propagator with mass $m_h$ 
summed to scalar propagator with a minus sign $(-i)$ and mass $M_{\phi}$. 
Thereby, the toy model (\ref{Leffh}) can be interpreted as one scalar theory of mass 
$m_{h}$ subtracted of another scalar theory with heavy mass of $M_{\phi}$.   
To check it explicitly, it is introduced the auxiliary scalar field $\tilde{h}$, 
such that the modified lagrangian is
\begin{eqnarray}\label{Lmodh}
{\cal L}_{sc}^{(\hat{h}\tilde{h})}=\frac{1}{2}\,(\partial_{\bar{\mu}}\hat{h})^2
-\frac{1}{2} \, m_{h}^2 \, \hat{h}^2 
-\tilde{h}\,\bar{\Box}\hat{h}
\nonumber \\
+\frac{1}{2}\,M_{\phi}^2\,\tilde{h}^2
-\lambda\,\sqrt{v}\,\hat{h}^3-\frac{\lambda}{4}\,\hat{h}^4 \, .
\;\;\;
\end{eqnarray}
The variation of (\ref{Lmodh}) in relation to $\tilde{h}$ yields the relation
\begin{eqnarray}
\tilde{h}=\frac{1}{M_{\phi}^2}\,\bar{\Box}\hat{h} \; ,
\end{eqnarray}
that when substituted in (\ref{Lmodh}), we recover the original lagrangian (\ref{Leffh}). 
We make the transformation $\hat{h}=h-\tilde{h}$ in (\ref{Lmodh}) to obtain the scalar lagrangian 
in terms of $h$ and $\tilde{h}$  
\begin{eqnarray}\label{Leffhh}
{\cal L}_{sc}^{(h\tilde{h})} &=& \frac{1}{2}\,(\partial_{\bar{\mu}}h)^2
-\frac{1}{2}\,(\partial_{\bar{\mu}}\tilde{h})^2
-\frac{1}{2} \, m_{h}^2 \, (h-\tilde{h})^2
\nonumber \\
&&
+\frac{1}{2} \, M_{\phi}^2 \, \tilde{h}^2
-\lambda\,\sqrt{v}\,(h-\tilde{h})^3-\frac{\lambda}{4}\,(h-\tilde{h})^4 \; .
\;\;\;\;
\end{eqnarray}
The previous lagrangian can be diagonalized by the transformation
\begin{subequations}
\begin{eqnarray}
h=h^{\prime} \, \cosh\alpha+\tilde{h}^{\prime} \, \sinh\alpha \; ,
\\
\tilde{h}=h^{\prime} \, \sinh\alpha+\tilde{h}^{\prime} \, \cosh\alpha \; , 
\end{eqnarray}
\end{subequations}
in which the $\alpha$-mixing angle satisfies the relation
\begin{eqnarray}
\tanh(2\alpha)=-\frac{2m_{h}^2}{M_{\phi}^2} \; .
\end{eqnarray}
For $M_{\phi} \gg m_{h}$, the mixing angle is very small, and the lagrangian (\ref{Leffhh}) 
in the basis of $h^{\prime}$ and $\tilde{h}^{\prime}$ is given by 
\begin{eqnarray}\label{Leffhhprime}
{\cal L}_{sc}^{(h^{\prime}\tilde{h}^{\prime})} &=& \frac{1}{2}\,(\partial_{\bar{\mu}}h^{\prime})^{2}
-\frac{1}{2} \, m_{h}^2 \, (h^{\prime})^2
\nonumber \\
&&
-\frac{1}{2}\,(\partial_{\bar{\mu}}\tilde{h}^{\prime})^2
+\frac{1}{2} \, M_{\phi}^2 \, (\tilde{h}^{\prime})^2
\nonumber \\
&&
-\lambda\,\sqrt{v}\,(h^{\prime}-\tilde{h}^{\prime})^3-\frac{\lambda}{4}\,(h^{\prime}-\tilde{h}^{\prime})^4 \; .
\end{eqnarray}
In (\ref{Leffhhprime}), we have two pseudo-scalars theories, the $h^{\prime}$-field sets a light scalar particle with mass 
$m_{h}$, and $\tilde{h}^{\prime}$ is a heavy scalar particle (that is the Lee-Wick scalar particle) 
with signs exchanged in the kinetic terms in relation to the terms of $h^{\prime}$. As consequence, 
the propagator of the $\tilde{h}^{\prime}$-field is
\begin{eqnarray}
D^{LW}(\bar{p})=\frac{-\,i}{\bar{p}^2-M_{\phi}^2} \; ,
\end{eqnarray}
that confirms the second term in the propagator (\ref{propPhisimpl}). The propagator (\ref{propPhi}) 
behaves like $\Delta(\bar{p}) \sim (\bar{p}^2)^{-2}$ in the ultraviolet regime, that leads to finite 
loop integrals for the scalar toy model (\ref{Leffh}) in $1+2$ dimensions. Thus, this scalar sector 
has finite contributions at the one-loop.
In the gauge sector, the massive term of $A^{\bar{\mu}}$ added to the gauge kinetic term of 
(\ref{PLLWresultOpN}) composes the pseudo Proca-Lee-Wick ED in 3D 
\begin{eqnarray} \label{PLLWresultafterSSB}
\mathcal{L}_{PLW} &=& -\frac{1}{4} \, F_{\bar{\mu}\bar{\nu}}N(\bar{\Box})F^{\bar{\mu}\bar{\nu}} +\frac{1}{2}\, (g^2v) \,A_{\bar{\mu}}^2
\nonumber \\
&&
+\frac{1}{2\xi} \, A_{\bar{\mu}}  \left(1+\frac{\bar{\Box}}{M^2} \right) N(\bar{\Box}) \, \partial^{\bar{\mu}}\,\partial^{\bar{\nu}} A_{\bar{\nu}} 
\; .
\end{eqnarray}
The propagator of the gauge sector after the SSB is
\begin{eqnarray}\label{propDelta}
\Delta_{\bar{\mu}\bar{\nu}}^{ssb}(\bar{k})&=&\frac{-1}{N(-\bar{k}^2)\,\bar{k}^2-m_a}\left[ \eta_{\bar{\mu}\bar{\nu}}-\frac{\bar{k}^2+M^2(\xi-1)}{\bar{k}^2-M^2}
\right.
\nonumber \\
&&
\left.
\times \frac{ N(-\bar{k}^2) \, k_{\bar{\mu}} \, k_{\bar{\nu}} }{ N(-\bar{k}^2)\,\bar{k}^2-\xi \, m_a  }
\right] \; ,
\end{eqnarray}
that in the ultraviolet range in which $\bar{k}^2 \gg \left\{ \, m_{a}^2 \, , \, M^{2} \, \right\}$, with $N(-\bar{k}^2) \simeq 1/\sqrt{-\bar{k}^2}$, the expression (\ref{propDelta}) 
goes to zero, and consequently, this behaviour corroborates to the renormalizability of the model in the perturbative analysis 
of a quantum field theory. Contracting (\ref{propDelta}) with the external conserved current $j^{\bar{\mu}}$, we obtain the amplitude :
\begin{eqnarray}\label{amp}
\mathcal{M}&=&{j^{\bar{\mu}}(\bar{k})\,\Delta^{ssb}_{\bar{\mu}\bar{\nu}}(\bar{k})\,j^{\bar{\nu}}}(\bar{k})
=\frac{- \, j_{\bar{\mu}}\,j^{\bar{\mu}}}{ N(-\bar{k}^2)\,\bar{k}^2-m_a }
\nonumber \\
&=&\frac{- \, j_{\bar{\mu}}\,j^{\bar{\mu}}\,[ \, N(-\bar{k}^2)\,\bar{k}^2+m_a \, ]}{ N^2(-\bar{k}^2)\,(\bar{k}^2)^2-m_a^2 } \; ,
\end{eqnarray}
in which the poles are defined by the solution of the equation
\begin{eqnarray}
N^2(-\bar{k}^2)\,(\bar{k}^2)^2-m_a^2=0 \; ,
\end{eqnarray}
whose roots are at $\bar{k}^{2}\simeq m_{a}^2$ and $\bar{k}^{2}\simeq M^{2}$, if $M \gg m_{a}$. Taking into account that $j^{\bar{\mu}}$ is space-like ($j^{2}<0$), 
see the refs. \cite{accioly03,accioly04,accioly05}, we have the residues
\begin{equation}\label{res}
\mbox{Res}\mathcal{M}(\bar{k}^{2}=m_{a}^2)>0  \; \; , \; \; \mbox{Res}\mathcal{M}(\bar{k}^{2}=M^{2})<0 \; .
\end{equation}
The above residues show that the model carries two spin-1 modes, one with a light mass at $\bar{k}^{2}=m_{a}^2$, and the other one with a heavier mass at $\bar{k}^{2}=M^2$. 
The positive residue corresponds to the Proca pseudo-electrodynamics, that describes a massive vector mode propagating non-locally on the plane. 
The negative residue is associated with the mode of a unstable heavy particle of mass $M$. 

%

\section{Some properties of the pseudo Proca-Lee-Wick electrodynamics}
\label{sec4}
\begin{figure}
\includegraphics[width=\linewidth]{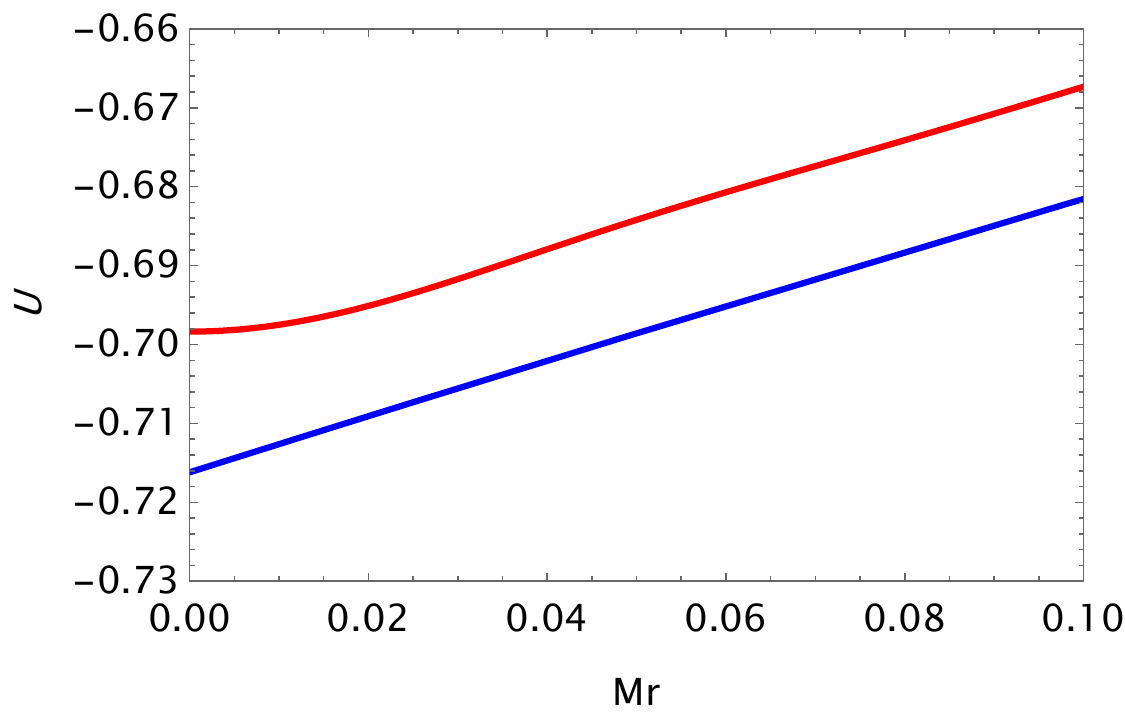}
\caption{The potentials as functions of dimensionless variable $Mr$. The blue line sets 
the Lee-Wick potential (\ref{UrLW}), and the red line is the numeric solution of (\ref{Urx}). 
The lines are illustrated for $e=0.3$ and $\mu_{a}=0.01$.} \label{Fig1}
\end{figure}

In this section, we study some classical features of the pseudo Proca-Lee-Wick electrodynamics. 
The Proca-Lee-Wick action associated with the lagrangian (\ref{PLLWresultafterSSB}) defined in the $1+2$ space-time in the presence of an external source $j^{\bar{\mu}}$ is
\begin{eqnarray}
S_{PLW}(A^{\bar{\mu}})=\int d^3\bar{x} \, \left[\,{\cal L}_{PLW} -j_{\bar{\mu}}\,A^{\bar{\mu}}\,\right] \; ,
\end{eqnarray}
in which the action principle yields the field equations
%
%
%
%
%
\begin{eqnarray}\label{EqX}
N(\bar{\Box})\,\partial_{\bar{\mu}}F^{\bar{\mu}\bar{\nu}}+m_{a}
A^{\bar{\nu}} = j^{\bar{\nu}} \; .
\end{eqnarray}
In our analysis here, we consider $\xi \rightarrow \infty$, such that the gauge fixing term can be removed.
The equations without sources are set by the Bianchi identity
\begin{eqnarray}
\partial_{\bar{\mu}}F_{\bar{\nu}\bar{\alpha}}+\partial_{\bar{\nu}}F_{\bar{\alpha}\bar{\mu}}+\partial_{\bar{\alpha}}F_{\bar{\mu}\bar{\nu}}=0 \; ,
\end{eqnarray}
that alternatively also can be represented by the divergence $\partial_{\bar{\mu}}\tilde{F}^{\bar{\mu}}=0$,
where $\tilde{F}^{\bar{\mu}}=\frac{1}{2}\,\epsilon^{\bar{\mu}\bar{\alpha}\bar{\beta}}F_{\bar{\alpha}\bar{\beta}}$ is the dual tensor of $F^{\bar{\mu}\bar{\nu}}$ 
in $1+2$ dimensions. The conservation current $\partial_{\bar{\mu}}j^{\bar{\mu}}=0$ leads to subsidiary condition for the $A^{\bar{\mu}}$-potential
$N(\bar{\Box})\,\partial_{\bar{\nu}}A^{\bar{\nu}} = 0$ in (\ref{EqX}). The components of the EM tensor provides the electric field on the
spatial plane, and the magnetic field perpendicular to this plane that we consider the ${\cal Z}$-direction, 
$F^{\bar{\mu}\bar{\nu}}=(E^{i},-\epsilon^{ij}\,B_{z})$, $i=\{\,1\,,\,2\,\}$. The matrix representation of $F^{\bar{\mu}\bar{\nu}}$ is
\begin{eqnarray}
F^{\bar{\mu}\bar{\nu}}=
\left(
\begin{array}{ccc}
0 & -E_{x} & -E_{y} \\
E_{x} & 0 & -B_{z} \\
E_{y} & B_{z} & 0 \\
\end{array}
\right) \; .
\end{eqnarray}
In vector notation, the equations of the pseudo-Proca-Lee-Wick electrodynamics are
\begin{subequations}
\begin{eqnarray}
N(\bar{\Box})(\nabla\cdot{\bf E})+m_{a}V &=& \sigma \; ,
\\
(\nabla \times {\bf E})_{z}+\partial_{t}B_{z} &=& 0 \; ,
\\
N(\bar{\Box})(\nabla\times{\bf B})+m_{a}{\bf A} &=& {\bf j}+N(\bar{\Box})\partial_{t}{\bf E} \; .
\end{eqnarray}
\end{subequations}
Using the subsidiary condition, the $A^{\bar{\mu}}$-potential satisfies the equation
\begin{eqnarray}
\left[ \, N(\bar{\Box})\,\bar{\Box} +m_{a} \, \right]A^{\bar{\nu}} = j^{\bar{\nu}} \; .
\end{eqnarray}
The conservation laws can be obtained directly from the field equations. Multiplying the equation (\ref{EqX}) by $F_{\bar{\nu}\bar{\alpha}}$, 
and using the Bianchi identity, we obtain the relation
\begin{eqnarray}\label{EqFJ}
\partial_{\bar{\mu}}\left[F_{\bar{\nu}\bar{\alpha}}\,N(\bar{\Box})\,F^{\bar{\mu}\bar{\nu}} \right]
+\partial_{\bar{\alpha}}\left[ \, \frac{1}{4} \, F_{\bar{\mu}\bar{\nu}}\,N(\bar{\Box})\,F^{\bar{\mu}\bar{\nu}} \, \right]
\nonumber \\
+\,m_{a} \, F_{\bar{\nu}\bar{\alpha}}\,A^{\bar{\nu}}=j^{\bar{\nu}}F_{\bar{\nu}\bar{\alpha}} \; .
\end{eqnarray}
The subsidiary condition allows to write the expression (\ref{EqFJ}) as the total derivative
\begin{eqnarray}\label{Eqtheta}
\partial_{\bar{\mu}}\theta^{\bar{\mu}}_{\;\;\,\bar{\alpha}}=j^{\bar{\nu}}\,F_{\bar{\nu}\bar{\alpha}} \; ,
\end{eqnarray}
where the energy-momentum tensor is given by
\begin{equation}
\theta^{\bar{\mu}}_{\;\;\,\bar{\alpha}}=F_{\bar{\nu}\bar{\alpha}}\,N(\bar{\Box})\,F^{\bar{\mu}\bar{\nu}}
+\,m_{a} \, A_{\bar{\alpha}}\,A^{\bar{\mu}}-\delta^{\bar{\mu}}_{\;\;\,\bar{\alpha}}\,{\cal L}_{PLW} \; .
\end{equation}
As we expect, the energy-momentum tensor is not gauge invariant when $m_{a} \neq 0$. The case without sources $j^{\nu}=0$ in (\ref{Eqtheta}) 
leads to conservation law $\partial_{\bar{\mu}}\theta^{\bar{\mu}}_{\;\;\,\bar{\alpha}}=0$, in which the conserved momentum for the Proca-Lee-Wick 
field is the integral of the $\theta^{0}_{\;\;\,\bar{\alpha}}$ - component over the spatial plane 
\begin{eqnarray}
P_{\bar{\alpha}}=\int d^2\bar{{\bf x}} \;  \theta^{0}_{\;\;\,\bar{\alpha}} \; .
\end{eqnarray}
The components $\bar{\alpha}=(0,i=1,2)$ yield the energy and the linear momentum densities stored in the Proca-Lee-Wick 
field, respectively,
\begin{subequations}
\begin{eqnarray}
\theta^{00}=\frac{1}{2} \, {\bf E}\,\cdot N(\bar{\Box}){\bf E}
+\frac{1}{2} \, B_{z}\,N(\bar{\Box})B_{z}
\nonumber \\
+\frac{1}{2} \, m_{a} (A_{0})^2
+\frac{1}{2} \, m_{a} \, {\bf A}^{2} \; ,
\\
\theta^{0i}=\epsilon^{ij}\left[N(\bar{\Box})\,E^{j} \right]B_{z}
-m_{a}\,A^{0} A^{i} \; .
\end{eqnarray}
\end{subequations}
For a static like point charge distribution, the current density is $j^{\bar{\mu}}(\bar{{\bf r}})=[-e \, \delta^{2}(\bar{{\bf r}}-{\bf r}),{\bf 0}]$,
and the effective action (\ref{actionEJ3D}) yields the static energy potential between two identical charges $(-e)$ separated by a distance $r$
\begin{equation}\label{Ur}
U(r)=-e^2 \int \frac{d^{2}\bar{{\bf k}}}{(2\pi)^2} \, \frac{e^{i\bar{{\bf k}}\cdot{\bf r}}}{ N(\bar{{\bf k}}^2)\,\bar{{\bf k}}^2+m_{a}} \, , \;\;
\end{equation}
that can be reduced to integral
\begin{equation}\label{Urx}
U(r)=-\frac{e^2\,M}{2\pi} \int_{0}^{\infty} dx \, \frac{x \, J_{0}(Mr \, x)}{2x(1+x^2+x\sqrt{1+x^2})+\mu_a} \; ,
\end{equation}
in which $\mu_a:=m_a/M$. This potential is plotted numerically as function of $Mr$ in comparison with the Lee-Wick potential (\ref{UrLW})
in the figure (\ref{Fig1}). We choose $e=0.3$ and $\mu_{a}=0.01$. The red line is the numerical solution of (\ref{Urx}) in that the potential
is finite at origin with $U(r=0)=-0.69$ in energy units. 

\section{The pseudo Proca-Lee-Wick quantum electrodynamics}
\label{sec5}
The addition of fermions to the Lagrangian density (\ref{PLLWresultafterSSB}) sets the Proca-Lee-Wick pseudo-quantum electrodynamics 
\begin{eqnarray}\label{LWPQED}
\mathcal{L}=\bar{\hat{\psi}}\left(i\,\slashed{D}-m\,\mathds{1}\right)\hat{\psi}+\frac{i}{M_{f}^2}\,\bar{\hat{\psi}} \, \slashed{D}\slashed{D}\slashed{D} \, \hat{\psi}
\nonumber \\
-\frac{1}{4}\,F_{\bar{\mu}\bar{\nu}}\,N(\bar{\Box})F^{\bar{\mu}\bar{\nu}}
+\frac{1}{2} \, m_{a} \, A_{\bar{\mu}}^2 \; ,
\end{eqnarray}
where $\slashed{D}=\Gamma^{\bar{\mu}}D_{\bar{\mu}}$ is the covariant derivative operator contracted with the Dirac matrices, 
$\hat{\psi}$ is a Dirac spinor (electron field) in $1+2$ dimensions, $m$ is the electron mass, $M_{f}$ is the Lee-Wick mass for the fermion,  
$\Gamma^{\bar{\mu}}=(\gamma^{0},\beta\,\gamma^{i})\,(i=1,2)$ are the Dirac matrices in which the spatial component includes 
the Fermi velocity $\beta \equiv v_F$, that satisfy the relation $\gamma^{\bar{\mu}}\gamma^{\bar{\nu}}=\eta^{\bar{\mu}\bar{\nu}}+i \, \sigma^{\bar{\mu}\bar{\nu}}$, 
and $\bar{\hat{\psi}}=\hat{\psi}^{\dagger}\gamma^{0}$ is the adjunct field. In the Dirac basis, the $\gamma^{0}-$ and $\gamma^{i}$-matrices are represented by
\begin{eqnarray}
\gamma^{0}&=&\sigma_{3}=
\left[
\begin{array}{cc}
1 & 0 \\
0 & -1 \\
\end{array}
\right]
\; , \;
\gamma^{1}=-i\sigma_3\,\sigma_{1}=\left[
\begin{array}{cc}
0 & -i \\
i & 0 \\
\end{array}
\right]
\; , \;
\nonumber \\
\gamma^{2}&=&-i\sigma_3\,\sigma_{2}=\left[
\begin{array}{cc}
0 & -1 \\
-1 & 0 \\
\end{array}
\right] \; ,
\end{eqnarray}
in which $\sigma_{i}=(\sigma_{1},\sigma_{2},\sigma_{3})$ are the Pauli matrices.
The covariant derivative operator couples the gauge field 
to the fermions as $D_{\bar{\mu}}=\partial_{\bar{\mu}}+i\,e\,A_{\bar{\mu}}$, in that the coupling constant $(e)$ is dimensionless 
in three dimensions (in the case of natural units $\hbar=c=1$). Taking the limits $M_{f} \rightarrow \infty$ and $M \rightarrow \infty$, 
the Lagrangian is reduced to the case of the pseudo-quantum ED in the presence of the mass $m_{a}$ \cite{Marinolivro}. 
The Dirac equation from (\ref{LWPQED}) coupled to the EM-field is
%
%
\begin{widetext}
\begin{eqnarray}
i\,\slashed{D}\left\{1+\frac{1}{M_{f}^2}\left[D_{0}^2-\beta^2\,\bar{D}_{i}^2+\frac{e\beta}{2}\left(
{\bm \sigma}\cdot{\bf E}-i({\bm \sigma}\times{\bf E})_{z}\right)+e\,\beta^2\,\sigma_{3}\,B_{z} \right]\right\}\hat{\psi}-m\,\hat{\psi}=0 \; .
\end{eqnarray}
\end{widetext}
In free sector of (\ref{LWPQED}), the fermion has the propagator in the momentum space
\begin{eqnarray}\label{propfermion}
S_{F}(p^{\bar{\mu}})=\frac{i}{\Gamma^{\bar{\mu}}p_{\bar{\mu}}\left(1- \bar{p}^2/M_{f}^2 \right)-m}
\nonumber \\
=\frac{i\,\left[\,\Gamma^{\bar{\mu}}p_{\bar{\mu}}\left(1-\bar{p}^2/M_{f}^2 \right)+m\,\right]}{\bar{p}^2
\left(1- \bar{p}^2/M_{f}^2 \right)^2-m^2} \; ,
\end{eqnarray}
in which $\bar{p}^2=p_{\bar{\mu}}\,p^{\bar{\mu}}=p_{0}^2-\beta^2\,{\bf p}^2$ for fermions in the momentum space.
The poles of (\ref{propfermion}) are determinate by the equation 
\begin{eqnarray}\label{polesfermions}
\bar{p}^2\left(\bar{p}^2-M_{f}^2 \right)^2-m^2\,M_{f}^{4}=0 \; ,
\end{eqnarray}
whose roots of are given by $\bar{p}^2 \simeq m^2$ and 
$\bar{p}^2 \simeq M_{f}^2$, if $M_{f} \gg m$. Therefore, 
the propagator (\ref{propfermion}) can be written as the combination
\begin{equation}\label{propcomb}
S_{F}(p_{0},p_{i})\simeq\frac{i}{\gamma^{0}p_{0}+\beta\gamma^{i}p_{i}-m}
-\frac{i}{\gamma^{0}p_{0}+\beta\gamma^{i}p_{i}-M_{f}} \; .
\end{equation}
The result (\ref{propcomb}) shows the combination of one propagator for a light fermion of mass 
$m$ subtracted of a propagator for the heavy Lee-Wick fermion of mass $M_{f}$. To see it explicitly 
in the lagrangian (\ref{LWPQED}), we introduce the auxiliaries fermion fields $\tilde{\psi}$ and 
$\tilde{\psi}^{\prime}$ such that the modified fermion lagrangian is 
\begin{eqnarray}\label{Lfpsis}
{\cal L}_{f}&=&\bar{\hat{\psi}}\left(i\,\slashed{D}-m\,\mathds{1}\right)\hat{\psi}+M_{\psi}\left( \bar{\tilde{\psi}}\,\tilde{\psi}^{\prime}+\bar{\tilde{\psi}}^{\prime}\,\tilde{\psi} \right)
\nonumber \\
&&
+\bar{\tilde{\psi}}\,i\,\slashed{D}\hat{\psi}+\bar{\hat{\psi}}\,i\,\slashed{D}\tilde{\psi}
-\bar{\tilde{\psi}}^{\prime}\,i\,\slashed{D}\tilde{\psi}^{\prime} \; ,
\end{eqnarray}
in which the action principle associated with $\bar{\tilde{\psi}}$ and $\bar{\tilde{\psi}}^{\prime}$ 
yield the relations, respectively,
\begin{eqnarray}
\tilde{\psi}^{\prime}=-\frac{i}{M_{f}}\,\slashed{D}\hat{\psi}
\; , \;
\tilde{\psi}=\frac{i}{M_{f}}\,\slashed{D}\tilde{\psi}^{\prime}
=\frac{1}{M_{f}^2}\,\slashed{D}\slashed{D}\hat{\psi} \; ,
\end{eqnarray}
that when substituted in (\ref{Lfpsis}), we recover the original fermion 
lagrangian from (\ref{LWPQED}). Making $\hat{\psi}=\psi-\tilde{\psi}$ in (\ref{Lfpsis}), 
the fermion lagrangian in terms of $\psi$, $\tilde{\psi}$ and $\tilde{\psi}^{\prime}$ is 
\begin{eqnarray}\label{Lfpsisresult}
{\cal L}_{f} &=& \bar{\psi}\left(i\,\slashed{D}-m\,\mathds{1}\right)\psi
+m\,\bar{\psi}\,\tilde{\psi}+m\,\bar{\tilde{\psi}}\,\psi-m\,\bar{\tilde{\psi}}\,\tilde{\psi}
\nonumber \\
&&
-\bar{\tilde{\psi}}\,i\slashed{D}\tilde{\psi}-\bar{\tilde{\psi}}^{\prime}\,i\slashed{D}\tilde{\psi}^{\prime}
+M_{f} ( \bar{\tilde{\psi}}\,\tilde{\psi}^{\prime}+\bar{\tilde{\psi}}^{\prime}\,\tilde{\psi} ) \; .
\end{eqnarray}
If we write $\tilde{\psi}$ and $\tilde{\psi}^{\prime}$ as the quiral components of a fermion $\chi$, 
$\tilde{\psi}=\chi_{L}=L\,\chi$ and $\tilde{\psi}^{\prime}=\chi_{R}=R\,\chi$, where $L=(\mathds{1}-\gamma_5)/2$ and $R=(\mathds{1}+\gamma_5)/2$ 
are the left- and right-handed projectors, in which $\gamma_{5}=i\,\gamma^{0}\gamma^{1}\gamma^{2}$, 
the second line in (\ref{Lfpsisresult}) reduces to the Lagrangian of the $\chi$-fermion 
with the Lee-Wick mass $M_{f}$ and the global minus sign, in which the mixed lagrangian with $\psi$ is
\begin{equation}
{\cal L}_{f}=\bar{\psi}\,i\,\slashed{D}-\bar{\chi}\,i\,\slashed{D}\chi 
-m\,\bar{\psi}\,\psi +m\,\bar{\psi}\,L\,\chi+m\,\bar{\chi}\,R\,\psi+M_{f}\,\bar{\chi}\,\chi \; .
\end{equation}
The lagrangian can be written in the matrix form
\begin{equation}\label{Lfupsilon}
{\cal L}_{f}=\bar{\Upsilon} \, K \,i\,\slashed{D}\Upsilon-\bar{\Upsilon} \, M_{\psi\chi} \, \Upsilon \; ,
\end{equation}
where $\Upsilon=( \, \psi \,\,\, \chi \, )^{t}$, and the kinetic matrix is $K=\mbox{diag}(1,-1)$, and the mass matrix $M_{\psi\chi}$ is
\begin{eqnarray}
M_{\psi\chi}=
\left[
\begin{array}{cc}
m & -L\, m \\
\\
-R\,m & -M_{f} \\
\end{array}
\right] \; .
\end{eqnarray}
The mass matrix in (\ref{Lfupsilon}) can be diagonalized by the transformation $\Upsilon^{\prime}=S\,\Upsilon$, 
in which $S$ is an unitary matrix, and $M_{D}=S\,M_{f}\,S^{\dagger}=\mbox{diag}(m,-M_{f})$ is the diagonal mass matrix 
with eigenvalues $m$ and $-M_{f}$, respectively. In the basis of $\psi^{\prime}$ and $\chi^{\prime}$, we obtain finally   
\begin{equation}
{\cal L}_{f}=\bar{\psi}^{\prime}\left(i\,\slashed{D}-m\,\mathds{1}\right)\psi^{\prime}
-\bar{\chi}^{\prime}\left(i\,\slashed{D}-M_{f}\,\mathds{1}\right)\chi^{\prime} \; .
\end{equation}
Physically, $\psi^{\prime}$ is a light fermion of mass $m$, and $\chi^{\prime}$ 
is the heavy Lee-Wick with mass $M_{f}$. The propagator of $\chi^{\prime}$ is
\begin{eqnarray}
S_{F}^{\chi^{\prime}}(p_{0},p_{i})=\frac{-\,i}{\gamma^{0}p_{0}+\beta\gamma^{i}p_{i}-M_{f}} \; ,
\end{eqnarray}
that explain the last term in (\ref{propcomb}). Therefore, the propagators of the fermion 
and of the Proca-Lee-Wick pseudo ED can help in renormalizability of the lagrangian (\ref{LWPQED}). 
Even if new couplings emerge from (\ref{LWPQED}), the investigation of the renormalization, 
or if the theory is finite in $1+2$ dimensions, is an important step for the consistency of a viable QFT.

%

%
%
%
%
%

%
\section{Unitarity in pseudo-Lee-Wick electrodynamics}
\label{sec6}
Unitarity is an important requirement for a quantum field theory. 
For systems in which the Hamiltonian operator does not depend on time explicitly, the operator 
$U(t,t_{0})=e^{-iH(t-t_0)}$ governs the time evolution of a state vector in quantum mechanics, 
where the unitarity imposes the condition $U^{\dagger}\,U=\mathds{1}$. The state vector 
is normalized in any time, and consequently, the probability is conserved. 
The unitarity of the $U$-operator implies the $S$-matrix is unitary, whose the elements are defined by
\begin{eqnarray}
S_{fi}=\lim_{t_{0} \rightarrow -\infty}\lim_{t \rightarrow +\infty}\langle f | e^{-iH(t-t_{0})} | i \rangle \; ,
\end{eqnarray}
where $|i\rangle$ and $|f\rangle$ are a complete set of free asymptotic states. The $S$-operator is written as 
$S=\mathds{1}+i\,T$, in which the unitarity $S^{\dagger}\,S=\mathds{1}$ leads to relation 
$i[T^{\dagger}-T]=T^{\dagger}\,T$, and defining the element $T_{ii}=\langle i | T | i \rangle$, the previous relation yields the result   
\begin{eqnarray}
2\,\mbox{Im}(T_{ii})=\sum_{f}T_{if}^{\dagger}\,T_{fi}=\sum_{f}T_{fi}^{\ast}\,T_{fi} \; ,
\end{eqnarray}
that is known as the Optical theorem. Writing $T_{ii}=(2\pi)^3\,\delta^3(0)\,\Delta_{F}(\bar{x}-\bar{x}^{\prime})$, in which 
$\Delta_{F}(\bar{x}-\bar{x}^{\prime})$ is the Feynman propagator for a scalar field, and the unitarity condition yields 
the relation
\begin{eqnarray}\label{RelDeltaF}
\Delta_{F}^{\ast}(\bar{x})-\Delta_{F}(\bar{x})=-i\int d\Phi \, (2\pi)^3\,\delta^{3}(0)
\nonumber \\
\times \int\frac{d^3\bar{x}^{\prime}}{(2\pi)^3} \, \Delta_{F}^{\ast}(\bar{x}^{\prime})
\,\Delta_{F}(\bar{x}-\bar{x}^{\prime}) \; ,
\end{eqnarray}
where $d\Phi$ is a phase factor needed to dimensional analysis that guarantee the sum over the 
intermediate states leads to identity. Using the Fourier transform in (\ref{RelDeltaF}), 
this relation in the momentum space is  
\begin{eqnarray}\label{GFrel}
G_{F}^{\ast}(\bar{k})-G_{F}(\bar{k})=-i \, {\cal T}^{-1} \, G_{F}^{\ast}(\bar{k})\,G_{F}(\bar{k}) \; ,
\end{eqnarray}
where ${\cal T}$ is the characteristic time of the system, that appears by the phase factor
\begin{eqnarray}
\int d\Phi \, (2\pi)^3\,\delta^{3}(0)={\cal T}^{-1} \; .
\end{eqnarray}
%
%
%

%
Since we are interested in the unitarity of a gauge theory at tree level, 
the relation (\ref{GFrel}) for a gauge field is modified by  
\begin{eqnarray}\label{GFrelgauge}
G_{F\bar{\mu}\bar{\nu}}^{\ast}(\bar{k})-G_{F\bar{\mu}\bar{\nu}}(\bar{k})=-i \, {\cal T}^{-1} \, G_{F\bar{\mu}\bar{\alpha}}^{\ast}(\bar{k})\,G_{F\bar{\alpha}\bar{\nu}}(\bar{k}) \; . \;\;\;\;
\end{eqnarray}
Therefore, the unitarity analysis starts with the Feynman propagator 
for the pseudo Lee-Wick electrodynamics (\ref{PLLWresultOpN}). 
The feynman propagator in the momentum space for (\ref{PLLWresultOpN}) 
in terms of the projectors $\theta_{\bar{\mu}\bar{\nu}}$ and $\omega_{\bar{\mu}\bar{\nu}}$ 
is 
\begin{eqnarray}\label{GFmunurel}
G_{F\bar{\mu}\bar{\nu}}(\bar{k})=A(\bar{k}^2)\,\theta_{\bar{\mu}\bar{\nu}}+B(\bar{k}^2)\,\omega_{\bar{\mu}\bar{\nu}} \; ,
\end{eqnarray}
where 
\begin{subequations}
\begin{eqnarray}
A(\bar{k}) &=& \frac{1}{2\sqrt{-\bar{k}^2-i\epsilon}}-\frac{1}{2\sqrt{-\bar{k}^2+M^2-i\epsilon}} \; ,
\label{Ak}
\;\;\;\;
\\
B(\bar{k}) &=& A(\bar{k}) \, \frac{M^{2}\,\xi}{M^2-\bar{k}^2} \; .
\end{eqnarray}
\end{subequations}
The term in $\omega_{\bar{\mu}\bar{\nu}}$ can be discarded by the gauge choice, or by the fact 
of the contraction with the conserved current, such that $j^{\bar{\mu}}\,\omega_{\bar{\mu}\bar{\nu}}\,j^{\bar{\nu}}=0$.
Thus, the Optical theorem (\ref{GFmunurel}) leads to relation
\begin{eqnarray}\label{ImAk}
2\,\Im[A(\bar{k})]=-{\cal T}^{-1}\,A^{\ast}(\bar{k})\,A(\bar{k}) \; .
\end{eqnarray}
Notice that $A(\bar{k})$ can be written as
\begin{eqnarray}\label{Akcomplex}
A(\bar{k}) = \frac{\sqrt{-\bar{k}^2+i\epsilon}}{2\sqrt{(\bar{k}^2)^2+\epsilon^2}}-\frac{\sqrt{-\bar{k}^2+M^2+i\epsilon}}{2\sqrt{(\bar{k}^2-M^2)^2+\epsilon^2}} \; , 
\end{eqnarray}
where the numerators are complex functions of $\bar{k}^2$, that are written in the polar form : 
$\chi_c:=\sqrt{-\bar{k}^2+i\epsilon}=\sqrt{\rho}\,e^{i\theta/2}$ and $\chi_M:=\sqrt{-\bar{k}^2+M^{2}+i\epsilon}=\sqrt{\rho_{M}}\,e^{i\alpha/2}$, 
in which $\rho=\sqrt{(\bar{k}^2)^2+\epsilon^2}$, $\rho_{M}=\sqrt{(\bar{k}^2-M^2)^2+\epsilon^2}$, with the arguments 
$\theta=\sin^{-1}(\epsilon/\rho)$ and $\alpha=\sin^{-1}(\epsilon/\rho_{M})$, respectively. Substituting (\ref{Akcomplex}) in (\ref{ImAk}), we obtain the relation 
\begin{equation}\label{EqChi}
\frac{\Im[\chi_{c}]}{\rho}-\frac{\Im[\chi_{M}]}{\rho_{M}}=-\frac{{\cal T}^{-1}}{4\rho\,\rho_{M}}\!\left[ \rho_{M}+\rho-2\sqrt{\rho_{M}\,\rho}\,\cos\left(\frac{\alpha-\theta}{2} \right) \right]
\end{equation}
Squaring the equation (\ref{EqChi}) and multiplying it by $\epsilon$, we use the identities 
\begin{subequations}
\begin{eqnarray}
\pi\,\delta(\bar{k}^2) &=& \frac{\epsilon}{(\bar{k}^2)^2+\epsilon^2} \; ,
\\
\pi\,\delta(\bar{k}^2-M^2) &=& \frac{\epsilon}{(\bar{k}^2-M^2)^2+\epsilon^2} \; ,
\end{eqnarray}
\end{subequations}
we obtain this relation evaluated at $\bar{k}^2=0$ and $\bar{k}^2=M^2$
\begin{eqnarray}\label{ImChik0}
&&
\left. \Im[\chi_c]^2 \right|_{\bar{k}^2=0}+\left. \Im[\chi_M]^2 \right|_{\bar{k}^2=M^2}
-\left. \frac{2\epsilon \, \Im[\chi_c] \, \Im[\chi_M] }{\sqrt{M^4+\epsilon^2}} \right|_{\bar{k}^2=0}
\nonumber \\
&&
=\frac{{\cal T}^{-2}}{16(M^4+\epsilon^2)}\!\!\left. \left[ \rho_{M}+\rho-2\sqrt{\rho_{M}\,\rho}\,\cos\left( \frac{\alpha-\theta}{2} \right) \right]^2 
\right|_{\bar{k}^2=0} \; .
\end{eqnarray}
Using the imaginary parts and the previous definitions, the expression (\ref{ImChik0}) yields  
relation of ${\cal T}$, $M$ and $\epsilon$ 
\begin{eqnarray}
\epsilon\left[1-\left(\frac{4}{1+M^2/\epsilon^2}\right)^{1/4}\!\!\sin\left(\frac{\beta}{2}\right) \right]=\frac{{\cal T}^{-2}}{16(M^4+\epsilon^2)} 
\nonumber \\
\times \left[\sqrt{M^4+\epsilon^2}+\epsilon-2\sqrt{\epsilon}(M^4+\epsilon^2)^{1/4}\cos\left(\frac{\beta}{2}-\frac{\pi}{4} \right) \right]^2 
, \;\;\;\;\;
\end{eqnarray}
where 
\begin{eqnarray}
\sin^2\left(\frac{\beta}{2}\right)=\frac{1}{2}\left[1-\frac{M^2}{\sqrt{M^4+\epsilon^2}} \right] \; .
\end{eqnarray}
For a small $\epsilon$-parameter, such that $M^2 \gg \epsilon$, $\sin\beta\simeq\epsilon/M^2$ and $\cos\beta\simeq 1$, that leads to relation
\begin{eqnarray}
{\cal T}^{-1}\simeq 4\sqrt{\epsilon} \left(1+\sqrt{\frac{2\epsilon}{M^2}} \right) \; .
\end{eqnarray}
This result is very similar to that obtained in ref. \cite{Marino2014} for the case of 
pseudo-quantum electrodynamics in the presence of radiative corrections at one-loop for the 
vacuum polarization. Therefore, we can confirm that the pseudo-Lee-Wick ED is unitary at tree level.

\section{Conclusions}
\label{sec7}
In this paper, we study the dimensional reduction of the Lee-Wick electrodynamics that leads to non-local
pseudo-Lee-Wick electrodynamics in $1+2$ dimensions, when the classical sources are confined on a spatial plane. 
The theory is a planar electrodynamics with derivatives of infinity order that is gauge invariant, and depends on 
the Lee-Wick mass parameter $(M)$, where the known Maxwell pseudo-electrodynamics is recovered in the limit $M \rightarrow \infty$.    
Thereby, this extend planar ED can mediate the interaction of electrons in a bidimensional material, in the presence of a massive 
gauge field that maintain the gauge invariance principle. As it is known, the Lee-Wick ED also present massless 
degree of freedom with propagation in the space-time, that similarly, keeps in the pseudo-Lee-Wick ED.  
Motived by planar superconductors models, a Higgs complex scalar field is introduced to yield mass like Proca 
to the massless sector of the pseudo-Lee-Wick ED, in which the $U(1)$ gauge symmetry is spontaneously broken in $1+2$ dimensions. 
Thus, one theory with a light massive degree freedom is obtained, 
beyond the heavy Lee-Wick mass. In this stage, the limit of $M \rightarrow \infty$ leads to a pseudo-Proca ED, where the Proca mass is 
acquired by a vacuum expected value scale. Some results, like the field equations and conservation laws are discussed on the 
classical pseudo-Proca-Lee-Wick ED. The sector of fermions in the Lee-Wick approach is introduced coupled to the 
Proca-Lee-Wick field. Thereby, the free sector contains a massive partner fermion that has a heavier mass in relation to the usual Dirac mass.
After some transformations, it is showed that the fermion Lagrangian is composite by a light Dirac fermion subtracted of a heavy Dirac fermion, 
both coupled to the Proca-Lee-Wick field. Posteriorly, the unitarity of the Lee-Wick pseudo ED is investigated through the Optical theorem at tree 
level. The non-local shape of the propagator shows that the theory is unitary in which the characteristic time is related to $\epsilon$-parameter from 
the Feynman propagator prescription. This result is similar to the case of the pure pseudo-ED.  
The proposal of the pseudo-Proca-Lee-Wick quantum electrodynamics opens the possibility of investigation for the unitarity of the theory 
in the presence of fermions, and taking account the radiative corrections. The renormalization of this theory also is an interesting investigation 
in $1+2$ dimensions since that the Proca-Lee-Wick and the fermion propagators have a good behavior in the ultraviolet regime. Furthermore, 
the loop integrals can have an improved divergency degree or finite by power counting in $1+2$ dimensions. Other perspective is studying the effects 
of the extended pseudo-QED in the presence of a topological Chern-Simons Lagrangian, that is like    
\begin{eqnarray}
{\cal L}_{CS}= i \, \frac{\lambda}{2} \, \epsilon^{\bar\mu\bar\nu\bar\rho} \, A_{\bar{\mu}} \, \partial_{\bar{\nu} }A_{\bar\rho}  \;  ,
\end{eqnarray}
where $\lambda$ is the dimensionless parameter, see the ref. \cite{Alves}. The radiative corrections can be calculated in one-loop approximation such 
that the electron self-energy, vacuum polarization and the vertex correction must be finite. 
The form factors and the electron anomalous magnetic moment can give a limit for the Lee-Wick mass parameter in the planar theory. 
These investigations are perspectives for a forthcoming project.

\end{document}